\documentclass[a4wide]{article}

\usepackage[T1]{fontenc}

\usepackage[bitstream-charter]{mathdesign}

\usepackage{graphicx}
\usepackage{bm}
\usepackage{stmaryrd}
\usepackage{mathtools}
    \mathtoolsset{showonlyrefs}
    \mathtoolsset{centercolon}

\renewcommand{\AA}{\ensuremath{\mathbf A}}
\newcommand{\BB}{\ensuremath{\mathbf B}}

\newcommand{\FF}{\ensuremath{\mathbf F}}
\newcommand{\HH}{\ensuremath{\mathbf H}}
\newcommand{\II}{\ensuremath{\mathbf I}}
\newcommand{\MM}{\ensuremath{\mathbf M}}
\newcommand{\PP}{\ensuremath{\mathbf P}}
\newcommand{\QQ}{\ensuremath{\mathbf Q}}

\newcommand{\TT}{\ensuremath{\mathbf T}}
\newcommand{\WW}{\ensuremath{\mathbf W}}

\renewcommand{\aa}{\ensuremath{\mathbf a}}
\newcommand{\bb}{\ensuremath{\mathbf b}}
\newcommand{\cc}{\ensuremath{\mathbf c}}
\newcommand{\ee}{\ensuremath{\mathbf e}}
\newcommand{\ff}{\ensuremath{\mathbf f}}
\renewcommand{\gg}{\ensuremath{\mathbf g}}
\newcommand{\hh}{\ensuremath{\mathbf h}}
\newcommand{\mm}{\ensuremath{\mathbf m}}
\newcommand{\nn}{\ensuremath{\mathbf n}}
\newcommand{\qq}{\ensuremath{\mathbf q}}
\renewcommand{\tt}{\ensuremath{\mathbf t}}
\newcommand{\vv}{\ensuremath{\mathbf v}}
\newcommand{\ww}{\ensuremath{\mathbf w}}
\newcommand{\xx}{\ensuremath{\mathbf x}}
\newcommand{\xxx}{\ensuremath{\bm{x}}}

\providecommand*{\ap}[1]{\ensuremath{^\mathrm{#1}}}

\newcommand{\divt}{\operatorname{div}}
\newcommand{\gradt}{\operatorname{grad}}
\newcommand{\curl}{\operatorname{curl}}
\newcommand{\secondgradient}{\gradt\negthinspace^2}
\newcommand{\sdiv}{{}^s\negmedspace\operatorname{div}}
\newcommand{\sgrad}{{}^s\negmedspace\operatorname{grad}}
\newcommand{\Hfil}{\mathsf{H}}
\newcommand{\Kfil}{\mathsf{K}}
\newcommand{\PPs}{{}^s\hspace{-.8pt}\PP}
\newcommand{\IIs}{{}^s\hspace{-.8pt}\II}

\newenvironment{remark}{\smallskip\noindent\emph{Remark.}}{\smallskip}

\begin{document}

\title{Hypertractions and hyperstresses convey the same mechanical information%
}


\author{Paolo Podio-Guidugli \\
              Dipartimento di Ingegneria Civile\\
    Universit\`a di Roma TorVergata \\
    Viale Politecnico 1, I-00133 Roma, Italy \\
              \texttt{email:~ppg@uniroma2.it} \medskip \\
    Maurizio Vianello \\
    Dipartimento di Matematica \\
    Politecnico di Milano\\
    Piazza Leonardo da Vinci 32, I-20133 Milano, Italy\\
              \texttt{email:~maurizio.vianello@polimi.it}
}

\maketitle

\begin{abstract}
A strengthened and generalized version of the standard Virtual Work Principle
is shown to imply, in addition to bulk and boundary balances, a one-to-one
correspondence between surface and edge hypertractions and hyperstrestress
fields in second-grade continua. When edge hypertractions are constitutively
taken null, the hyperstress is shown to take the form it has for a
Navier-Stokes$-\alpha$ fluid, a relevant example of second-grade fluid-like
material.
\end{abstract}

\section{Introduction}
    \label{Sec:1}

The main conceptual point we want to make in this note is that stipulating a
suitable Principle of Virtual Powers to characterize mechanical equilibrium of
continua of any grade bigger than one offers a key advantage: a Cauchy-type
construction of the hyperstress fields accompanying the equilibrium
hypertraction fields (a difficult task, that has been undertaken but not
achieved so far)  is no more needed, because \emph{hyperstresses can be
explicitly computed in terms of hypertractions, and conversely}. In this paper,
we demonstrate this tenet in the case of second-gradient continua, by a simple
argument.

Theories of second-gradient continua have a long history. In the case of
fluids, a possible dependence of pressure on the density gradient was first
proposed by Korteweg~\cite{Korteweg1901} to model capillarity effects in 1901;
for solids, two pioneering papers by Toupin~\cite{Toupin1962,Toupin1964}  on
elastic materials with couple stresses appeared in the early years 1960. A PVP
approach to the formulation of the basic balance laws for these continua was
taken by Germain~\cite{Germain1972,Germain1973,Germain1973a} in the early
1970s; the comprehensive article by Maugin~\cite{Maugin1980} appeared in 1980;
a recent contribution, of special relevance to our present paper, is due to
Gurtin and Fried~\cite{Fried2006}.

It is well known that compatibility with the second law of thermodynamics for
constitutive relations which are functions of the second deformation gradient
demands that internal mechanical interactions have a nonstandard form. This has
been discussed by many authors, mainly with an eye towards a better
clarification of the role of the so-called hypertractions and hyperstresses; an
important contribution was given by Dunn and Serrin~\cite{Dunn1985}, who
introduced the notion of the interstitial energy. An interesting feature of
second-gradient materials is that, if bodies and subbodies having non
everywhere smooth boundary are considered, then {\em edge forces}, that is,
line distributions of hypertractions are to be expected (and, if a dependence
on gradients higher than two is allowed, one has to deal also with {\em vertex
forces}, as exemplified by Podio-Guidugli~\cite{Podio-Guidugli2002}). To our
knowledge, a rigorous interaction theory accommodating such a nonstandard
behavior remains to be constructed; interesting attempts in this direction have
been carried out by Forte and Vianello~\cite{FV88}, Noll and
Virga~\cite{Noll1990}, and Dell'Isola and Seppecher~\cite{DS1997a}.

Although here we do not deal with this difficult issue directly, in Section
\ref{Sec:3}, the bulk of this paper, we do provide a full set of
\emph{representation formulae} not only, as is relatively easy, for tractions
and hypertractions, both diffused and concentrated on edges, in terms of
stresses and hyperstresses (see definitions \eqref{hatt}$_1$, \eqref{hatt}$_2$,
and \eqref{hatef}) but also, conversely, \emph{for stresses and hyperstresses
in terms of diffused and concentrated tractions and hypertractions} (see
\eqref{repTtilde}-\eqref{repThat}, and \eqref{hatHfil}). Such representation
formulae generalize the corresponding formulae for simple ($\equiv$
first-gradient) materials, that we derive in our preparatory Section
\ref{Sec:2}. Since we work in a nonvariational setting, our results apply
whatever the material response. The PVP we use includes edge tractions, both
internal and external; without them, it would not be possible to arrive at the
complete representation formula for the hyperstress in terms of hypertractions
we construct in Subsection~\ref{hh}.

Finally, in Section~\ref{Sec:4}, we provide a new proof of the following not
very well-known fact in the theory of second-gradient materials: if edge
tractions are constitutively presumed null on whatever edge, then the
hyperstress needs not be zero, although it takes a very special form whose
information content is carried by a vector field. We surmise that inability to
develop edge interactions be characteristic of certain \emph{second-gradient
fluids}, an issue that we take up in a forthcoming paper
\cite{Podio-Guidugli2009}, continuing a line of thought proposed in
\cite{Podio-Guidugli2007}.

\section{Simple Continua}\label{Sec:2}

\subsection{Power expenditures as constitutive requirements}

When a characterization of mechanical equilibrium is sought via a weak
formulation of the virtual-work type, the primary object is a linear space
$\mathcal V$ of \emph{test}($\equiv\,$\emph{virtual}) \emph{velocity fields};
the collection of \emph{tractions} is introduced as the formal dual of
$\mathcal V$, by laying down a notion of external power expended in a virtual
body motion; and the collection of \emph{stresses} is introduced as the formal
dual of the collection of test-velocity gradients, by laying down a notion of
internal power. We regard specification of  these two duality relations  as the
`zeroth grade' of any constitutive theory.

We classify a material body $B$ as \emph{simple} if the \emph{internal} and
\emph{external power expenditures} have the following forms:
\begin{equation}\label{mvw}
\begin{aligned}
    {\mathcal W}^{(i)}(P)[\vv]&:=\int_{P}\TT\cdot\gradt\vv,\\
    {\mathcal W}^{(e)}(P)[\vv]&:=\int_{\partial P}\tt\cdot\vv\,,
\end{aligned}
\end{equation}
for all \emph{body parts}($\equiv \,$\emph{subbodies}) $P$ and for all
test-velocity fields $\vv\in{\mathcal V}$. Here $P$ is a bounded subset of the
current observation space being \emph{regularly open} (that is, coinciding with
the interior of its closure) and having a \emph{part-wise regular boundary
$\partial P$}; we regard it as the region occupied by the typical body part at
the time of our observation; time itself plays the role of a parameter. As to
$\mathcal V$, we accept the standard assumption that it includes all realizable
velocities (that is, all velocity fields obtained by time differentiation of
admissible deformation fields) and that it is \emph{closed under the operation}
${\mathcal O}$ \emph{of observer change}, in the sense that, if
$\vv\in{\mathcal V}$, then
\begin{equation}\label{velO}
\vv^+={\mathcal O}(\vv)=\dot\qq+\QQ\vv+\WW\xx^+\in{\mathcal V}
\end{equation}
for all translation velocities $\dot\qq$, all rotations ($\equiv$ proper
orthogonal tensors) $\QQ$, and all relative spins ($\equiv$ skew-symmetric
tensors) $\WW$ (for details, see the first subsection of the Appendix).

The  \emph{internal} field $\TT$, the \emph{Cauchy stress}, is meant to measure
the mechanical interaction of a material element of the subbody $P$ with its
immediate adjacencies; the \emph{external} field $\tt$, the \emph{contact
traction}, is meant to account for the mechanical action exerted on $P$ by its
complement with respect to the material universe the body $B$ belongs to. (We
intentionally ignore all types of actions at a distance, because they are
inessential to the purpose of our present discussion.)

Both fields $\TT$ and $\tt$ are here introduced formally by way of Riesz
duality with, respectively, the test velocity fields $\vv$ and their gradient
fields $\gradt\vv$. \emph{Both power expenditures $\TT\cdot\gradt\vv$ and
$\tt\cdot\vv$ are required to be properly invariant under observer changes}.
Precisely, translational invariance of the external power expenditure over an
arbitrary subbody implies that \emph{the contact traction field be balanced},
i.e., that
\begin{equation}
\int_{\partial P}\tt={\mathbf 0}\quad \mbox{for all subbodies $P$.}
\end{equation}
Moreover, the specific external power expenditure is rotationally invariant if
and only if \emph{the contact traction is indifferent to observer changes}, in
the sense that
\begin{equation}\label{iot}
    {\mathcal O}(\tt)=\tt^+=\QQ\tt\quad\mbox{for all rotations $\QQ$}.
\end{equation}

As to the specific internal power expenditure, which is quickly seen to be
invariant under translational observer changes, its rotational invariance
implies that, at all points of $B$, \emph{the Cauchy stress field be
symmetric-valued and indifferent to observer changes}, in the sense that
\begin{equation}\label{ioT}
  {\mathcal O}(\TT)= \TT^+=\QQ\TT\QQ^T
\end{equation}
(for a proof of this result, see the Appendix).

\subsection{Mutual consistency of stresses and tractions via a strengthened
Principle of Virtual Powers}

The mutual consistency of the stress and traction fields is the consequence of
postulating the following \emph{Principle of Virtual Power}:
\begin{equation}\label{PVP}
{\mathcal W}^{(i)}(P)[\vv]={\mathcal W}^{(e)}(P)[\vv],
\end{equation}
\emph{for all body parts} $P$ and \emph{for all test-velocity fields} $\vv$.

Needless to say, the Principle is an invariant statement. The quantification on
velocities is standard, that on body parts is not. Asking that (\refeq{PVP})
holds for all body parts is much stronger a requirement than demanding it to
hold only for the whole body.\footnote{We have been unable to assess who
introduced this strenghtened qunatification in continuum mechanics first, when
and where. Needless to say, without it, it would not be possible to
characterize equilibrium for a system of rigid bodies, nor the method of Euler
cuts would make any sense.} This additional strength connects the values taken
by $\TT$ and $\tt$ at  all points of $B$, and not only at its boundary points.

With the use of a standard integration-by-parts lemma, we find that
\begin{equation}
    \int_{P}\TT\cdot\gradt\vv
    =\int_{P}(-\divt\TT)\cdot\vv+\int_{\partial P}\TT\nn\cdot\vv.
\end{equation}
Consequently, \eqref{PVP} can be written as follows:
\begin{equation}\label{pP}
    \int_{P}(-\divt\TT)\cdot\vv+\int_{\partial P}\TT\nn\cdot\vv
    =\int_{\partial P}\tt\cdot\vv,
\end{equation}
for all body parts $P$ and for all test velocity fields $\vv$; in particular,
\begin{equation}\label{pB}
    \int_{B}(-\divt\TT)\cdot\vv+\int_{\partial B}\TT\nn\cdot\vv
    =\int_{\partial B}\tt\cdot\vv,
\end{equation}
for all test velocity fields $\vv$. Now, while both (\refeq{pP}) and
(\refeq{pB}) imply the classic pointwise balance:
\begin{equation}
    -\divt\TT={\mathbf 0}\quad \mbox{at all points of $B$,}
\end{equation}
it is (\refeq{pP}) that implies that
\begin{equation}\label{cons}
    \TT\nn=\tt\quad \mbox{at all points of $B$ and for all unit vectors $\nn$,}
\end{equation}
and not only at the points of $\partial B$ and for $\nn$ the outward unit
normal field, as implied by the weaker statement (\refeq{pB}).

A first direct consequence of (\refeq{cons}) is that
\begin{equation}
    \tt=\hat\tt(x,\nn),
\end{equation}
namely, that -- within the class of simple continua -- the traction field at
any point $x\in B$ must be thought of as  depending on the orientation of the
plane chosen through that point in order to detect the mutual mechanical
contact interactions. Secondly, it follows from (\refeq{cons}) that the fields
$\TT$ and $\tt$ carry essentially the same information. Indeed, while, given
the Cauchy stress mapping $x\mapsto\hat\TT(x)$, (\refeq{cons}) yields that
\begin{equation}\label{trazrep}
\hat\tt(x,\nn)=\hat\TT(x)\nn,
\end{equation}
we also have that, given the traction mapping $(x,\nn)\mapsto\hat\tt(x,\nn)$,
the Cauchy stress field can be constructed as follows:
\begin{equation}\label{stresrep}
    \hat\TT(x)=\sum_i\hat\tt(x,\nn^{(i)})\otimes\nn^{(i)},
\end{equation}
for $\nn^{(i)}\;(i=1,2,3)$ any three mutually orthogonal directions. Thus,
(\refeq{cons}) can be regarded as a basic \emph{consistency condition} for the
pair of dynamic quantities $(\tt,\TT)$ dual to the kinematic quantities
$(\vv,\gradt\vv)$, a condition that yields the \emph{representation formulae}
(\refeq{trazrep}) and (\refeq{stresrep}).

All these results  are more or less well known in the continuum mechanics
community. The main reason for us to recapitulate them is that they prompt the
following remarks, that are at the core of our present work.

\subsection{Thinking of complex continua}\label{comcont}

Relations (\refeq{trazrep}) and (\refeq{stresrep}) are also arrived at when, as
is customary, only tractions on body parts are introduced, because stress is
constructed \emph{\`a la} Cauchy as a consequence of balance of
tetrahedron-shaped parts. The Cauchy construction is the pillar on top of which
the standard theory of \emph{diffuse} (i.e., absolutely continuous with respect
to the area measure) contact interactions stands.  For  \emph{complex} (i.e.,
nonsimple) material bodies, a Cauchy-like construction has been attempted
often, but not achieved so far, to our knowledge. Consequently, for such
continua, although there is some sort of a general agreement about what notion
of diffuse contact interactions applies, it is not clear what generalized
stresses should accompany them to achieve mechanical balance.  Now, as we just
showed for simple material bodies acted upon by diffuse contact interactions, a
key advantage of a Principle of Virtual Power quantified over a large
collection of body parts is that it dispenses us from going through Cauchy's
tetrahedron construction to represent stress in terms of tractions. In the next
section, this feature of  a virtual-power approach to formulate mechanical
balance is demonstrated in the case of a popular type of complex continua: we
show what consistency relations restrict the choices of
tractions/hypertractions and stresses/hyperstresses, and we derive
representation formulae that generalize relations (\refeq{trazrep}) and
(\refeq{stresrep}), where edge hypertractions have a crucial role.

\section{Second-Gradient Continua}\label{Sec:3}

\subsection{Notation}

$\mathsf H$, a capital letter from a sansserif font, is used to denote the
third-order hyperstress tensor. We write $H_{ijk}$ for the cartesian components
of $\mathsf H$ with respect to a fixed orthonormal basis ${\mathbf e}_i$. For
any vector $\aa$, we write $\mathsf H\aa$ for the second-order tensor with
components $(\Hfil\aa)_{ij}=H_{ijk}a_k$ (summation over repeated indexes
understood); and,  for any second-order tensor $\AA$, we write $\mathsf H[\AA]$
for the vector whose cartesian components are $(\Hfil[\AA])_i=H_{ijk}A_{jk}$;
in particular, $\mathsf H[\aa\otimes\bb]=(\mathsf H\bb)\aa$.

Let $\mathcal V$ be the translation space associated to the euclidean space
where $\mathcal S$, a smooth surface, is embedded. Given any point of $\mathcal
S$, we denote by $\PPs$ the orthogonal projection of the vector space $\mathcal
V$ onto the tangent space to $\mathcal S$ at that point.  The transpose of
$\PPs$ is $\IIs$, the inclusion map, which takes any vector in the tangent
space into a vector of  $\mathcal V$. Thus, for $\nn$ a unit normal to the
surface and $\II$ the identity mapping over $\mathcal V$,
$\IIs\PPs=\II-\nn\otimes\nn$.\footnote{We need not use here any of the
well-known representations of $\PPs$ and $\IIs$. The interested reader is
referred to a paper by Gurtin and Murdoch~\cite{Gurtin1974}.}

\subsection{Virtual power expenditures}

Among complex material bodies, \emph{second-gradient continua} are those for
which the internal and external power expenditures have the following forms:
\begin{itemize}
\item (\emph{internal power expenditure})
\begin{equation}\label{vwg1}
    {\mathcal W}^{(i)}(P)[\vv]:=
    \int_{P}\TT\cdot\gradt\vv+\Hfil\cdot\secondgradient\vv,
\end{equation}
where $\secondgradient\,\vv:=\gradt(\gradt\vv)$ is the second spatial
gradient of the velocity field, and $\Hfil$ is a third-order tensor field
which satisfies:
\begin{equation}\label{simacca}
    (\Hfil\aa)\bb=(\Hfil\bb)\aa,\quad \textrm{for all vectors $\aa,\bb$},
\end{equation}
so as to have the same index symmetries as the field $\secondgradient\,\vv$
it is dual to;
\item (\emph{external power expenditure})
\begin{equation}\label{vwg2}
    {\mathcal W}^{(e)}(P)[\vv]:=
    \int_{\partial P}\big(\tt\cdot\vv+\hh\cdot\partial_\nn\vv\big)
    +\int_{\widehat{\partial P}}\ff\,\ap{e}\cdot\vv\,,
\end{equation}
where $\widehat{\partial P}$ is the `edgy' part of $\partial P$, if any
(once again, and for the same reasons as before, any distance interaction
entering the external power has been ignored).
\end{itemize}

These definitions require some comments. Firstly, we here consider a part
collection larger than usual: the boundary $\partial P$ of a part may happen to
be the union of a finite number of regular surfaces which join pairwise along
\emph{edges}, that is, smooth curves which begin and end at singular points
called \emph{vertices}; accordingly, the edgy part $\widehat{\partial P}$ of
$\partial P$ is the union of a finite set of boundary curves where the
surface-normal field has a jump.\footnote{A more precise and formal description
of these geometrical notions is found in a paper by Noll \&
Virga~\cite{Noll1990}.} Secondly, the introduction in \eqref{vwg1} of the
additional internal power expenditure associated with the \emph{hyperstress}
$\Hfil$ is paraleled by the introduction in \eqref{vwg2} of two types of
external power expenditures associated with \emph{hypertractions}, both
additional to the one associated with the traction field $\tt$, namely, the
\emph{diffused hypertraction} field $\hh$ and the \emph{edge-force} field
$\ff\,\ap{e}$. Thirdly, the consequences of requiring that both power
expenditures be properly invariant under observer changes must be investigated.
We treat the two last issues in the following subsections.

\subsubsection{Diffused and concentrated hypertractions}

Suppose that, for consistency with postulating an additional power expenditure
in the bulk associated with the second velocity gradient, one associates with
the first the following additional power expenditure at a part's boundary:
\begin{equation}
    \int_{\partial P}\HH\cdot\gradt\vv,
\end{equation}
with $\HH$ some second-order tensor field work-conjugated to $\gradt\vv$. Then,
on splitting the velocity gradient into its tangential and normal parts:
\begin{equation}
\gradt\vv=(\sgrad\vv)\PPs+\partial_\nn\vv\otimes\nn,
\end{equation}
we have that
\begin{equation}\label{formula}
    \int_{\partial P}\HH\cdot\gradt\vv
    =\int_{\partial P}(\HH\IIs)\cdot\sgrad\vv
    +\int_{\partial P}\HH\nn\cdot\partial_\nn\vv\,.
\end{equation}
Now, on setting $\HH\nn=\hh$, the second integral on the right side of
(\refeq{formula}) can be identified with the second surface integral in
(\refeq{vwg2}); to motivate the presence of the line integral, we need a
consequence of a surface-divergence identity, that we introduce right away.

Let $\mathcal S$ be  a regular surface, oriented by its normal $\nn$. At each
point $\xxx$ of its boundary curve $\partial {\mathcal S}$, a unit vector $\mm$
can be chosen, orthogonal to both $\nn$ and the tangent direction of $\partial
{\mathcal S}$ and pointing outward from the interior of $\mathcal S$; such a
vector $\mm$ lies in the limiting tangent plane to $\mathcal S$ at $\xxx$, and
can be represented as  $\mm=\IIs\bm{\mu}$, with  $\bm{\mu}$ the appropriate
tangent vector. For any smooth tensor field $\AA$ over $\mathcal S$, the
following integral identity holds true:
\begin{equation}\label{identity}
    \int_{\mathcal S} (\AA\IIs)\cdot\sgrad \vv
    =-\int_{\mathcal S}\sdiv(\AA\IIs)\cdot\vv
    +\int_{\partial{\mathcal S}}\AA\mm\cdot\vv\,.
\end{equation}

\begin{figure}
    \centering
  \includegraphics{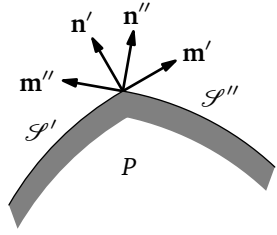}
    \caption{Four unit vectors describe an edge of a body part.}
    \label{fig:generic.edge}
\end{figure}

With a view to applying this identity to each regular portion of the boundary
of an edgy body part, we draw attention to Figure~\ref{fig:generic.edge},
that depicts an edge cross-section of a part $P$.  The limiting values of the
outward unit normals to the surfaces $\mathcal S'$ and $\mathcal S''$ joining
at the edge are $\nn'$ and $\nn''$, while the unit vectors $\mm'$ and $\mm''$
belong to the tangent planes to those surfaces, and point outward from their
interior. The edge $\mathcal E$ is described by an ordered list of four unit
vectors:
\begin{equation}\label{edge}
    \mathcal E\equiv(\nn',\mm';\nn'',\mm'')\,.
\end{equation}
Note that the vectors in the list are not independent, because they are
coplanar and pairwise orthogonal; moreover, the ordered pairs $(\nn',\mm')$ and
$(\nn'',\mm'')$ can be interchanged freely. With the stipulated conventions,
(\refeq{edge}) gives a local (first-order) description of the geometry of an
edge, just as a unit normal provides a local (first-order) description of an
oriented surface.

With this notation, the identity (\refeq{identity}) yields:
\begin{equation}\label{iddiv}
    \int_{\partial P}(\HH \IIs)\cdot\sgrad\vv =
    - \int_{\partial P}\sdiv(\HH\IIs)\cdot\vv
    +\int_{\widehat{\partial P}}\llbracket\HH\mm\rrbracket\cdot\vv,
\end{equation}
where $\llbracket\HH\mm\rrbracket$ denotes twice the \emph{edge average} of
$\HH\mm$:
\begin{equation}
    \llbracket\HH\mm\rrbracket:=\HH\mm' + \HH\mm'',
\end{equation}
a vector field over $\widehat{\partial P}$. Thus, while the first in
(\refeq{iddiv}) can be safely thought as absorbed into the first surface
integral in (\refeq{vwg2}), the second motivates the introduction in
(\refeq{vwg2}) itself of the line integral over the edgy part of ${\partial
P}$.

\subsubsection{Invariance of power expenditures under observer changes}

Translational and rotational invariances of the external power expenditure over
an arbitrary subbody imply, respectively, the \emph{hypertraction balance}
\begin{equation}
\int_{\partial P}\tt\;+\int_{\widehat{\partial P}}\ff\,\ap{e}
={\mathbf 0}\quad \mbox{for all subbodies $P$.}
\end{equation}
and the \emph{indifference properties}
\begin{equation}\label{iott}
   \tt^+=\QQ\tt,\quad \hh^+=\QQ\hh,\quad ({\ff\,\ap{e}})^+=
   \QQ\ff\,\ap{e}\quad\mbox{for all rotations $\QQ$}.
\end{equation}
It is shown in the Appendix that the rotational invariance of the specific
internal power expenditure implies that, at all points of $B$, (i) \emph{the
stress field $\TT$ be symmetric-valued and indifferent to observer changes},
just as it was the case for the Cauchy stress field in simple continua, and
(ii) \emph{the hyperstress field $\Hfil$ be indifferent to observer changes},
in the sense that
\begin{equation}\label{ioTT}
  \Hfil^+= \QQ\ast\Hfil,\quad\textrm{with}\;\;(\QQ\ast\Hfil)[\aa\otimes\bb]\cdot\cc
  =\Hfil[\QQ^T\aa\otimes\QQ^T\bb]\cdot\QQ^T\cc\,.
\end{equation}

\subsection{The Principle of Virtual Powers and its consequences}

We postulate the following \emph{Principle of Virtual Powers}:
\begin{equation}\label{PVPP}
{\mathcal W}^{(i)}(P)[\vv]={\mathcal W}^{(e)}(P)[\vv],
\end{equation}
\emph{for all body parts} $P$ and \emph{for all test velocity fields} $\vv$.

We want to show that there are two types of consequences of the part-wise
balance principle (\refeq{PVPP}):
\begin{itemize}
\item[\textit{i}.] point-wise \emph{bulk and boundary balances}, the same
    that would follow from a standard global Principle, stated for $P\equiv
    B$ only and quantified over the collection of test velocity fields;
\item[\textit{ii}.] \emph{mutual consistency relations} for the stress
    fields $(\TT,\Hfil)$ and the traction fields $(\tt,\hh,\ff\,\ap{e})$,
    such that either list determines uniquely the other, thus allowing the
    conclusion that each list conveys the same mechanical information.
\end{itemize}

We begin by noticing that, via repeated integration by parts, we have that
\begin{equation}
\begin{aligned}
    &\displaystyle{\int_{P}\TT\cdot\gradt\vv+\Hfil\cdot(\secondgradient\vv)}
    =\displaystyle{\int_{P}\big(\TT-\divt\Hfil)\cdot\gradt\vv
    + \int_{\partial P}\Hfil\nn\cdot\gradt\vv} \\
    =&\displaystyle{\int_P(-\divt\widetilde\TT)\cdot\vv +
    \int_{\partial P}}\big(\widetilde\TT\nn\cdot\vv
    +(\Hfil\nn)\nn\cdot\partial_\nn\vv+(\Hfil\nn)\cdot\sgrad\vv\big),
\end{aligned}
\end{equation}
where we have set
\begin{equation}\label{titilde}
\widetilde\TT:=\TT-\divt\Hfil.
\end{equation}
Next, just as we deduced (\refeq{iddiv}) from (\refeq{identity}), we find that
\begin{equation}
    \int_{\partial P}(\Hfil\nn)\cdot\sgrad\vv
    = - \int_{\partial P}\sdiv((\Hfil\nn)\IIs)\cdot\vv
    +\int_{\widehat{\partial P}}\llbracket(\Hfil\nn)\mm\rrbracket\cdot\vv.
\end{equation}
Thus, the PVP relation (\refeq{PVPP}) can be given the following provisional
form:
\begin{equation}\label{prevp}
\begin{gathered}
    \int_P(-\divt\widetilde\TT)\cdot\vv+\int_{\partial P}
    \Big(\big(\widetilde\TT\nn-\sdiv((\Hfil\nn)\IIs)\big)\cdot\vv
    +(\Hfil\nn)\nn\cdot\partial_\nn\vv\Big) \\
    +\int_{\widehat{\partial P}}
    \llbracket(\Hfil\nn)\mm\rrbracket\cdot\vv
    = \int_{\partial P}\big(\tt\cdot\vv
    +\hh\cdot\partial_\nn \vv\big)+\int_{\widehat{\partial P}}\ff\,\ap{e}\cdot\vv\,,
\end{gathered}
\end{equation}
for all subbodies and all virtual velocity fields. 	

By exploiting both quantifications \emph{with the use of smooth body parts
only}, we have from (\ref{prevp}) that, \emph{at all points of $B$}, the
following balances must hold:
\begin{equation}\label{1}
-\divt\widetilde\TT={\mathbf 0},
\end{equation}
with $\widetilde\TT$ defined by \eqref{titilde} in terms of $\TT$ and $\Hfil$;
moreover, \emph{for each unit vector} $\nn$,
\begin{equation}\label{2}
    \widetilde\TT\nn-\sdiv((\Hfil\nn)\IIs) =
    \tt,\quad\textrm{for all smooth surfaces oriented by}\;\nn;
\end{equation}
and
\begin{equation}\label{2bis}
(\Hfil\nn)\nn = \hh.
\end{equation}
Notice that the last relation can also be written as
\begin{equation}\label{hyperstress.on.surface}
\Hfil[\nn\otimes\nn]=\hh,\quad\textrm{for all unit vectors}\;\nn.
\end{equation}

The remain of (\refeq{prevp}) is
\begin{equation}
    \int_{\widehat{\partial P}}\big(\llbracket(\Hfil\nn)\mm\rrbracket
    -\ff\,\ap{e}\big)\cdot\vv=0\,,
\end{equation}
\emph{for all edgy body parts} and all test velocity fields. Consequently, we
have that, again \emph{at all points of $B$},
\begin{equation}\label{f.at.edge}
    \llbracket(\Hfil\nn)\mm\rrbracket=
    \ff\,\ap{e},\quad\textrm{for all edges}\;{\mathcal E},
 \end{equation}
or rather, more explicitly,
\begin{equation}\label{f.at.edge.bis}
  \Hfil[\nn'\otimes\mm'+\nn''\otimes\mm'']=\ff\,\ap{e},\quad\textrm{for all edges as in Figure~\ref{fig:generic.edge}}.
\end{equation}

\begin{remark}
Differences in notation apart, equations \eqref{1}, \eqref{2}-\eqref{2bis} and
\eqref{f.at.edge} coincide respectively with equations (4), (5) and (49) of
Gurtin and Fried~\cite{Fried2006}; as they remark, when applied to the body's
boundary, those equations are the same as equations (7.8), (7.9) and (7.10),
derived variationally by Toupin in \cite{Toupin1962}. Our equation \eqref{2},
however, is written in much more compact a form than the corresponding
equations in \cite{Fried2006} and \cite{Toupin1962}. Interestingly, as noticed
earlier by Forte and Vianello~\cite{FV88}, this opens the way to a suggestive
analogy with
\begin{equation}
\FF\nn-\sdiv\MM=\tt,
\end{equation}
the balance equation in \emph{shell theory} that involves the \emph{force and
moment tensors} $\FF$ and $\MM$, that is, the surface-stress descriptors of
that theory; the following identifications suffice:
\begin{equation}
\FF\equiv\widetilde\TT,\quad \MM\equiv(\Hfil\nn)\IIs.
\end{equation}
\end{remark}

\subsection{Bulk and boundary balances}

Relation (\refeq{1}) is nothing but the
\begin{itemize}
   \item{(\emph{bulk balance\/})}
\begin{equation}\label{pwisebal}
    -\divt\widetilde\TT={\mathbf 0}\quad\textrm{ in $B$}\,.
\end{equation}
Relations \eqref{2} and \eqref{2bis} have, as we are going to see, a manifold
use.

Firstly, let $(\partial B)\ap{reg}_{\ff}$ denote the portion of the body's
regular boundary
\begin{equation}
  (\partial B)\ap{reg}:=\partial B\setminus\widehat{\partial B}
\end{equation}
where the external force and hyperforce fields $\tt_0$ and $\hh_0$ are
assigned. Then, by a standard pill-box argument, we can deduce from
\eqref{2} the
    \medskip
  \item{(\emph{diffused traction \& hypertraction boundary balances\/})}
\begin{equation}\label{diffbc}
    \widetilde\TT\nn-\sdiv((\Hfil\nn)\IIs)
    =\tt_0,\quad (\Hfil\nn)\nn=\hh_0\quad\mbox{over $(\partial B)\ap{reg}_{\ff}$.}
\end{equation}
Moreover, on $(\widehat{\partial B})_{\ff}$, the portion of the edgy part
of $\partial B$, if any, where external line forces $\ff\,\ap{e}_0$ are
assigned, we deduce from (\refeq{f.at.edge}) the\medskip
    \item{(\emph{concentrated hypertraction boundary balance\/})}
\begin{equation}\label{concbc}
    \llbracket(\Hfil\nn)\mm\rrbracket
    =\ff\,\ap{e}_0\quad\mbox{over $(\widehat{\partial B})_{\ff}$.}
\end{equation}
\end{itemize}

\subsection{Representation formulae}\label{hh}

Given the stress and hyperstress fields $\widehat\TT$ and $\widehat{\mathsf H}$
over $B\cup\partial B$, relations \eqref{titilde}, \eqref{2}, and \eqref{2bis},
can be used to define the \emph{traction and hypertraction mappings}:
\begin{equation}\label{hatt}
\begin{aligned}
    \hat\tt_{\mathcal S}(x,\nn)&:=
    \big(\widehat\TT(x)-\divt\widehat\Hfil(x)\big)\nn-\sdiv(\widehat\Hfil(x)\nn)\IIs,\\
    \hat\hh(x,\nn)&:=\big(\widehat\Hfil(x)\nn\big)\nn,
\end{aligned}\end{equation}
for each unit vector $\nn$ and all smooth surfaces $\mathcal S$ oriented by
$\nn$; likewise, relation  \eqref{f.at.edge.bis} can be used to define the
\emph{line-traction mapping} $\ff\,\ap{e}=\hat{\ff\,\ap{e}}(x,{\mathcal E})$:
\begin{equation}\label{hatef}
    \hat{\ff\,\ap{e}}(x,{\mathcal E}):=
    \widehat\Hfil(x)[\nn^\prime\otimes\mm^\prime+
    \nn^{\prime\prime}\otimes\mm^{\prime\prime}],
    \quad
    \mathcal E\equiv(\nn',\mm';\nn'',\mm''),
\end{equation}
for all edges $\mathcal E$, both of the body and of its parts.

Our next goal is the deduction of a representation formula for
$\widehat{\mathsf{H}}(x)$  in terms of the diffused and concentrated
hypertractions acting, respectively, on three oriented coordinate planes
through $x$ and the relative coordinate edges; with such a formula at hand, it
is easy to deduce a representation formula for $\widetilde\TT(x)$ that
resembles \eqref{stresrep}):
\begin{equation}\label{repTtilde}
    \widetilde\TT(x)=\sum_i\big(\hat\tt(x,\nn_i)
    +\sdiv(\widehat{\mathsf H}(x)\nn_i\IIs)\big)\otimes\nn_i,
\end{equation}
whence, for $\widehat\TT(x)$:
\begin{equation}\label{repThat}
\widehat\TT(x)=\widetilde\TT(x)+\divt\widehat{\mathsf H}(x).
\end{equation}

\begin{remark}
In addition to the complications that we are going to display along our path to
determine $\widehat{\mathsf H}$ in terms of the surface and edge hypertraction
mappings $\hat\hh$ and $\hat{\ff\,\ap{e}}$,  the complicated dependence
\eqref{repTtilde} of $\widetilde\TT$ on the diffused traction mapping $\hat\tt$
and on $\widehat{\mathsf H}$ itself gives a first idea of the difficulties
intrinsic to any attempt to generalize Cauchy's tetrahedron construction we
alluded at in Subsection \ref{comcont} to second-gradient materials (let alone
materials of higher grade).
\end{remark}

\begin{figure}
    \centering
  \includegraphics{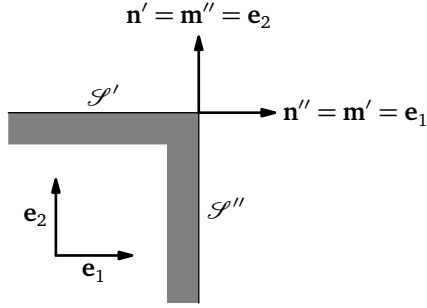}
    \caption{The coordinate edge $\mathcal E_{12}$.}
    \label{fig:coordinate.edge.e1e2}
\end{figure}

\begin{figure}
    \centering
  \includegraphics{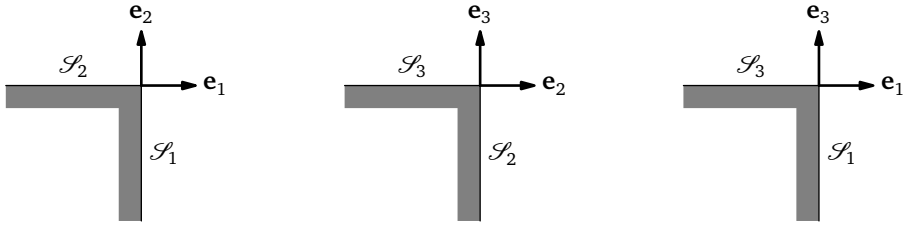}
    \caption{The coordinate edges $\mathcal E_{12}$, $\mathcal E_{23}$, $\mathcal E_{13}$.}
    \label{fig:coordinate.edges.v2}
\end{figure}

At a given body point, let $\mathcal S_1$ be the \emph{oriented plane} parallel
to $\ee_2, \ee_3$ whose normal unit vector is $\ee_1$, and let the oriented
planes $\mathcal S_2$ and $\mathcal S_3$ be similarly defined. The
\emph{coordinate edge} $\mathcal E_{12}$ is depicted in
Figure~\ref{fig:coordinate.edge.e1e2};
it is constructed as a special case of the edge in
Figure~\ref{fig:generic.edge}, by taking $\mathcal S_1$ and $\mathcal S_2$ for
$\mathcal S''$ and $\mathcal S'$, respectively, and by choosing $\mm'=\ee_2$,
$\mm''=\ee_1$ (such prescriptions are needed to identify $\mathcal E_{12}$
unequivocally); the coordinate edges $\mathcal E_{23}$ and $\mathcal E_{13}$
are similarly defined (Figure~\ref{fig:coordinate.edges.v2}).

With the use of \eqref{hyperstress.on.surface} and \eqref{f.at.edge.bis} we
define, respectively, the diffuse hypertractions $\hh_i$ on the coordinate
planes $\mathcal S_i$ and the concentrated forces $\ff\,\ap{e}_{ik}$ on the
coordinate edges $\mathcal E_{ik}$, namely,
\begin{equation}\label{expressions.for.hi}
    \begin{aligned}
  \hh_1 = \mathsf H[\ee_1\otimes\ee_1], \\
    \hh_2 = \mathsf H[\ee_2\otimes\ee_2], \\
      \hh_3 = \mathsf H[\ee_3\otimes\ee_3], \\
\end{aligned}
\end{equation}
and
\begin{equation}\label{expressions.for.ff.jk}
    \begin{aligned}
  \ff\,\ap{e}_{12} &=\mathsf H[\ee_1\otimes\ee_2+\ee_2\otimes\ee_1]=2\, \mathsf H[\ee_1\otimes\ee_2], \\
    \ff\,\ap{e}_{13} &=\mathsf H[\ee_1\otimes\ee_3+\ee_3\otimes\ee_1]=2\, \mathsf H[\ee_1\otimes\ee_3], \\
      \ff\,\ap{e}_{23} &= \mathsf H[\ee_2\otimes\ee_3+\ee_3\otimes\ee_2]= 2\, \mathsf H[\ee_2\otimes\ee_3], \\
\end{aligned}
\end{equation}
where we have accounted for the required symmetry of $\mathsf H$, expressed by
\eqref{simacca}. We now show that the six vectors $\hh_i$ and
$\ff\,\ap{e}_{ik}$ (each of which does depend on the Cartesian basis chosen)
allow for the construction of a basis-independent representation formula for
the hyperstress tensor $\mathsf H$.

Indeed, we know that
\begin{equation}
  \mathsf H= \sum_{ijk} H_{ijk} \ee_i\otimes \ee_j\otimes \ee_k;
\end{equation}
again taking the index symmetry of $H_{ijk}$ into account, this sum can be
rewritten as
\begin{equation}\label{sum.for.H.rewritten}
\begin{aligned}
\mathsf H &=
   \sum_j \sum_i H_{ijj} \ee_i\otimes \ee_j\otimes \ee_j +
   \sum_{j\ne k} \sum_i H_{ijk} \ee_i\otimes \ee_j \otimes \ee_k  \\
     &= \sum_j \sum_i H_{ijj} \ee_i\otimes \ee_j\otimes \ee_j +
   \sum_{j<k} \sum_i  H_{ijk} \ee_i\otimes (\ee_j \otimes \ee_k +\ee_k \otimes \ee_j). \\
 \end{aligned}
\end{equation}
Since
\begin{equation}
  H_{ijk}=\ee_i\cdot\mathsf H[\ee_j\otimes\ee_k]
\end{equation}
we deduce that
\begin{equation}
  H_{ijk}=\frac12 \ee_i\cdot\mathsf H[\ee_j\otimes\ee_k+\ee_k\otimes\ee_j].
\end{equation}
Thus, in view of \eqref{expressions.for.hi},
\begin{equation}
H_{ijj} = \frac12 \ee_i\cdot\mathsf H[\ee_j\otimes\ee_j+\ee_j\otimes\ee_j] = \frac12 \ee_i \cdot  2 \hh_j = \ee_i \cdot \hh_j,
\end{equation}
which implies that
\begin{equation}\label{acca}
  \hh_j =\hat\hh(x,\ee_j)= \sum_i \widehat{H}_{ijj}(x)\ee_i \quad (\mbox{no sum over $j$}).
\end{equation}
Similarly, for $j<k$,
\begin{equation}
    H_{ijk} = \frac12 \ee_i\cdot\mathsf H[\ee_j\otimes\ee_k+\ee_k\otimes\ee_j] = \frac12 \ee_i\cdot \ff\,\ap{e}_{jk},
\end{equation}
which implies that
\begin{equation}\label{effe}
   \ff\,\ap{e}_{jk} = \hat\ff\,\ap{e}(x,{\mathcal E}_{jk})=2 \sum_i \widehat{H}_{ijk}(x)\ee_i, \quad (\mbox{for $j<k$}).
\end{equation}
By substitution of \eqref{acca} and \eqref{effe} in
\eqref{sum.for.H.rewritten}, we deduce the desired representation formula for
$\widehat{\mathsf H}$ in terms of $\hat\hh$ and $\hat\ff\,\ap{e}$:
\begin{equation}\label{hatHfil}
\widehat{\mathsf H}(x) = \sum_j \hat\hh(x,\ee_j) \otimes \ee_j\otimes \ee_j +
    \frac12 \sum_{j<k}  \hat\ff\,\ap{e}(x,{\mathcal E}_{jk}) \otimes (\ee_j \otimes \ee_k+\ee_k \otimes \ee_j).
\end{equation}
We regard this formula as the main result of this paper.

\section{Second-gradient materials with no edge tractions}
    \label{Sec:4}

We now ask the following question: which restrictions on $\mathsf H$ can be
deduced from the assumption that \emph{\textmd{no forces act on any edge, real
or imagined}}? Said differently: is the hyperstress  necessarily everywhere
null in a second-gradient material body deemed to be unable to develop edge
tractions?

Let us then suppose that, at a fixed body point $x$, we have that
$\hat\ff\,\ap{e}(x,\cdot)\equiv\bm{0}$, so that the forces on all coordinate
edges are zero, whatever the basis. Under this assumption, we have from
\eqref{hatHfil} that, with respect to any given basis, $\mathsf H$  can be
writte as:
\begin{equation}\label{H.no.edge.forces}
  \mathsf H= \sum_j \hh_j \otimes \ee_j\otimes \ee_j
\end{equation}
(recall that vectors $\hh_j$ do depend on the chosen basis). Now, consider a
new orthonormal basis $\bar{\ee}_i$, obtained from $\ee_i$ by a rotation about
$\ee_3$ of some angle $\theta$:
\begin{equation}\label{e.bar.e}
  \bar{\ee}_1=\cos\theta\,{\ee}_1+\sin\theta\,{\ee}_2,\quad
    \bar{\ee}_2=-\sin\theta\,{\ee}_1+\cos\theta\,{\ee}_2,\quad
    \bar{\ee}_3={\ee}_3,
\end{equation}
and compute $${\bar{\ff}}\,\ap{e}_{12}=2 \mathsf H[\bar{\ee}_1\otimes
\bar{\ee}_2].$$ It turns out that
\begin{equation}
\begin{aligned}
   {\bar{\ff}}\,\ap{e}_{12} &= -2\sin\theta\cos\theta\,
 \underbrace{\mathsf H[{\ee}_1\otimes{\ee}_1]}_{=\,\hh_1}
  +2\sin\theta\cos\theta\, \underbrace{\mathsf H[{\ee}_2\otimes{\ee}_2]}_{=\,\hh_2} \\
  &\quad\quad +\cos^2\theta\,\underbrace{2\mathsf H[{\ee}_1\otimes{\ee}_2]}_{=\,{{\ff}}\,\ap{e}_{12}}
  -\sin^2\theta\,\underbrace{2\mathsf H[{\ee}_2\otimes{\ee}_1]}_{=\,{{\ff}}\,\ap{e}_{12}}\,.
\end{aligned}
\end{equation}
But, because the edge tractions ${\bar{\ff}}\,\ap{e}_{12}$ and
${{\ff}}\,\ap{e}_{12}$ must be zero by assumption, it follows from the last
relation that, for all $\theta$,
\begin{equation}
  \bm{0}=2\sin\theta\cos\theta (\hh_2-\hh_1);
\end{equation}
this implies that $\hh_1=\hh_2$. Moreover, for a rotation about $\ee_1$, the
same argument  yields that $\hh_2=\hh_3$. Then, we can set
\begin{equation}
  \hh:=\hh_1=\hh_2=\hh_3,
\end{equation}
and write \eqref{H.no.edge.forces} as follows:
\begin{equation}
  \mathsf H = \hh \otimes \sum_j \ee_j\otimes\ee_j
    = \hh \otimes \mathbf I.
\end{equation}
We conclude that a constitutive lack of edge forces implies that
\begin{equation}\label{Hh}
\mathsf H=\hh\otimes \mathbf I,
\end{equation}
for some vector $\hh$; such $\hh$ does not depend on the basis chosen for the
argument, because $3\hh=\mathsf H[\mathbf I]$. Conversely, if $\mathsf H$
admits the representation \eqref{Hh} for some vector $\hh$, then it readily
follows from \eqref{f.at.edge.bis} and \eqref{edge} that all edge tractions are
zero. Indeed,
\begin{equation}
\begin{aligned}
  \hat\ff\,\ap{e}(\mathcal E) &=
   \mathsf H[\nn'\otimes\mm'+\nn''\otimes\mm''] \\
   &= \big(\mathbf I\cdot(\nn'\otimes\mm'+\nn''\otimes\mm'')\big)\,\hh \\
   &= (\nn'\cdot\mm'+\nn''\cdot\mm'')\,\hh =\bm{0},
\end{aligned}
\end{equation}
since $(\nn',\mm')$ and $(\nn'',\mm'')$ are pairs of orthogonal vectors.

\begin{remark} An alternative proof of the assertion leading to \eqref{Hh}
was given by Forte and Vianello in \cite{FV88}. Here it is. If there are no
edge tractions, then
\begin{equation}
\ww\cdot \mathsf H[\nn\otimes \mm]=0
\end{equation}
for any given vector $\ww$ and for all pairs of orthonormal vectors $\nn$,
$\mm$.  Thus, the second-order symmetric tensor ${\mathsf H}^T\ww$, whose
components are  $({\mathsf H}^T\ww)_{jk}=w_i H_{ijk})$, must have diagonal form
with respect to any basis:
\begin{equation}
  \ww\mathsf H=\lambda(\ww)\mathbf I,\quad\mbox{for some scalar-valued function
$\lambda(\ww)$}.
\end{equation}
However, since the left side of this relation is linear in $\ww$, there must be
a vector $\hh$ such that $\lambda(\ww)=\hh\cdot\ww$. But,
\begin{equation}
  \ww\mathsf H=(\hh\cdot\ww)\, \mathbf I\quad \Rightarrow\quad  \mathsf H=\hh\otimes \mathbf I\,.
\end{equation}
\end{remark}

\begin{remark}
No mathematically precise notion of fluidity and solidity has been suggested so
far for complex continua. A current malpractice is to base the distinction on
whether the first-gradient stress arising in response to deformation histories
complies with the one or the other of the group-theoretic recipes proposed by
Noll to sort simple fluids from simple solids. Recently, one of us
\cite{Podio-Guidugli2007} came up with a notion of macroscopic aggregation
state of matter based on an invariance requirement of the internal power
expenditure under classes of changes of the reference placement different for
fluid and solid simple materials. This notion is formally easy to generalize to
complex materials, although the analysis involved is far from trivial. We
develop these issues in full in a forthcoming paper \cite{Podio-Guidugli2009}.
Suffice it to mention here a class of \emph{Navier-Stokes$-\alpha$ fluids},
that is, of second-gradient materials obeying, among others to be introduced
shortly, the following constitutive prescription for the power expenditure of
hypertractions:
\begin{equation}
\mathsf H\cdot\secondgradient\vv=\gg\cdot\curl\curl\vv\,.
\end{equation}

In view of the differential identity:
\begin{equation}
\curl\curl\vv=\gradt(\divt\vv)-\Delta\vv,
\end{equation}
it is not difficult to see that
\begin{equation}
\gg\cdot\curl\curl\vv=\Hfil_1\cdot\secondgradient\vv+\Hfil_2\cdot\secondgradient\vv;
\end{equation}
with
\begin{equation}
(\Hfil_1)_{ijk}=\frac 1 2\big(\delta_{ij}g_k+\delta_{ik}g_j\big)\quad\textrm{and}\quad \Hfil_2=-\gg\otimes\II.
\end{equation}
Furthermore, if
\begin{equation}
\divt\vv=0,
\end{equation}
as is customarily stipulated in microfluidics, then
\begin{equation}
\gg\cdot\curl\curl\vv=-\gg\cdot\Delta\vv.
\end{equation}
Thus, (i) $\Hfil_1$ can be interpreted as the \emph{reactive hyperstress}
associated to the second-gradient consequence of the standard
\emph{incompressibility constraint}; (ii) $\Hfil_2$, the \emph{active
hyperstress}, induces no edge hypertractions; and, (iii) if one chooses the
simplest constitutive prescription
\begin{equation}
\gg=\zeta\,\Delta\vv,\quad\zeta>0,
\end{equation}
then the regularizing term $-\zeta\,\Delta(\Delta\vv)$ typical of the
Navier-Stokes$-\alpha$ flow equation \cite{Fried2006} appears in the balance
equation.
\end{remark}

\section{Appendix}

We collect here some subsidiary material, with the purpose of improving the
readability and self-containment of our paper.

\subsection{Observer changes}
By an \emph{observer change} we mean the following transformation rule for
position vectors of space points:
\begin{equation}\label{oc}
  \xx^+={\mathcal O}(\xx)=\qq+\QQ\xx,\quad \xx:=x-o,
\end{equation}
where $\qq$ is an arbitrary time-dependent translation and $\QQ$ an arbitrary
time-depen\-dent rotation  about point $o$. The related transformation rule for
velocities is:
\begin{equation}
  \vv^+=\dot\qq+\QQ\vv+\WW\xx^+,\quad \WW:=\dot\QQ\QQ^T=-\WW^T,
\end{equation}
where all of the parameters $\dot\qq$, $\QQ$, and $\WW$, can take arbitrary
values at any chosen time (cf. \eqref{velO}). It follows from \eqref{oc} that
\begin{equation}\label{gradpiu}
\gradt^+(\cdot)=\gradt(\cdot)\QQ^T\,.
\end{equation}
Hence, as to the velocity gradient, one finds that
\begin{equation}\label{gradiente}
  (\gradt\vv)^+=\QQ(\gradt\vv)\QQ^T+\WW.
\end{equation}

\subsection{Rotational invariance of the internal power expenditure}

The rotational invariance of the specific internal power expenditure is the
requirement that, at all points of $B$,
\begin{equation}\label{second.power.invariance}
  \TT\cdot\gradt\vv+\Hfil\cdot\secondgradient\vv=
    \TT^+\cdot(\gradt\vv)^+ +\Hfil^+\cdot(\secondgradient\vv)^+,
\end{equation}
where
\begin{equation}
\TT^+={\mathcal O}(\TT),\quad \Hfil^+={\mathcal O}(\Hfil), \quad\textrm{etc.}
\end{equation}

In order to discuss the consequences of this requirement, it is convenient to
define the \emph{action of the rotation group on second- and third-order
tensors}: in cartesian components,
\begin{equation}
  [\QQ\ast\AA]_{ij}:=Q_{ip}Q_{jq}A_{pq},\qquad
    [\QQ\ast\Hfil]_{ijk}:=Q_{ip}Q_{jq}Q_{kr}H_{pqr};
\end{equation}
in absolute notation,
\begin{equation}
  \QQ\ast\AA=\QQ\AA\QQ^T,
  \qquad
  (\QQ\ast\Hfil)[\aa\otimes\bb]\cdot\cc
  =\Hfil[\QQ^T\aa\otimes\QQ^T\bb]\cdot\QQ^T\cc\,.
\end{equation}
Notice that, for any pair of rotations $\QQ_1$ and $\QQ_2$,
\begin{equation}
  (\QQ_1\QQ_2)\ast\AA=\QQ_1\ast(\QQ_2\ast\AA),
  \qquad
    (\QQ_1\QQ_2)\ast\Hfil=\QQ_1\ast(\QQ_2\ast\Hfil);
\end{equation}
moreover,
\begin{equation}\label{action.transpose.scalar}
  (\QQ\ast\AA)\cdot\BB=\AA\cdot(\QQ^T\ast\BB),
  \qquad
  (\QQ\ast\Hfil)\cdot\Kfil=\Hfil\cdot(\QQ^T\ast\Kfil),
\end{equation}
for all pairs of second- and third-order tensors $\AA$, $\BB$, $\Hfil$, $\Kfil$
and for all rotations $\QQ$.

With this notation, the transformation rule \eqref{gradiente} becomes:
\begin{equation}\label{gradi}
  (\gradt\vv)^+=\QQ\ast(\gradt\vv)+\WW.
\end{equation}
Moreover, by an application of \eqref{gradpiu}, we readily deduce the
transformation rule for the second gradient of velocity:
\begin{equation}
  (\secondgradient\vv)^+=\QQ\ast(\secondgradient\vv).
\end{equation}
Thus, \eqref{second.power.invariance} can be rewritten as:
\begin{equation}\label{second.power.invariance.bis}
  \TT\cdot\gradt\vv+\Hfil\cdot\secondgradient\vv=
    \TT^+\cdot(\QQ\ast\gradt\vv+\WW)+\Hfil^+\cdot(\QQ\ast\secondgradient\vv),
\end{equation}
or rather, equivalently in view of \eqref{action.transpose.scalar},
\begin{equation}
  \TT\cdot\gradt\vv+\Hfil\cdot\secondgradient\vv=
    (\QQ^T\ast\TT^+)\cdot\gradt\vv+\TT^+\cdot\WW
    +(\QQ^T\ast\Hfil^+)\cdot\secondgradient\vv.
\end{equation}
Since all of the multipliers $\gradt\vv$, $\secondgradient\vv$ and $\WW$ can be
prescribed arbitrarily and independently from each other, we conclude that
$\TT$ must be symmetric and, moreover, that the following transformation rules
must hold:
\begin{equation}
  \TT^+ =\QQ\ast\TT,\qquad \Hfil^+ =\QQ\ast\Hfil
\end{equation}
(cf.~\eqref{ioTT}). This conclusion fully generalizes the well known result for
simple (first order) materials.



\bibliographystyle{spmpsci}

\bibliography{HandH}   

\begin{thebibliography}{10}
\providecommand{\url}[1]{{#1}}
\providecommand{\urlprefix}{URL }
\expandafter\ifx\csname urlstyle\endcsname\relax
  \providecommand{\doi}[1]{DOI~\discretionary{}{}{}#1}\else
  \providecommand{\doi}{DOI~\discretionary{}{}{}\begingroup
  \urlstyle{rm}\Url}\fi

\bibitem{DS1997a}
Dell'Isola, F., Seppecher, P.: Edge contact forces and quasi-balanced power.
\newblock Meccanica \textbf{32}, 33--52 (1997)

\bibitem{Dunn1985}
Dunn, J.E., Serrin, J.: On the thermodynamics of interstitial working.
\newblock Arch. Rational Mech. Anal. \textbf{88}, 95--133 (1985)

\bibitem{FV88}
Forte, S., Vianello, M.: On surfaces stresses and edge forces.
\newblock Rend. Mat. Appl. \textbf{8}(3), 409--426 (1988)

\bibitem{Fried2006}
Fried, E., Gurtin, M.E.: Tractions, balances, and boundary conditions for
  nonsimple materials with application to liquid flow at small-length scales.
\newblock Arch. Rational Mech. Anal. \textbf{{182}}({3}), {513--554} ({2006}).
\newblock \doi{{10.1007/s00205-006-0015-7}}

\bibitem{Germain1972}
Germain, P.: Sur l'application de la m\'ethode des puissances virtuelles en
  m\'ecanique des milieux continus.
\newblock C. R. Acad. Sc. Paris S\'erie A \textbf{274}, 1051--1055 (1972)

\bibitem{Germain1973}
Germain, P.: La m\'ethode des puissances virtuelles en m\'ecanique des milieux
  continus. premi\`ere partie: Th\'eorie du second gradient.
\newblock J. M\'ecanique \textbf{12}, 235--274 (1973)

\bibitem{Germain1973a}
Germain, P.: The method of virtual power in continuum mechanics. part 2:
  Microstructure.
\newblock SIAM J. Appl. Math. \textbf{25}, 556--575 (1973)

\bibitem{Gurtin1974}
Gurtin, M.E., Murdoch, I.: A continuum theory of elastic material surfaces.
\newblock Arch. Rational Mech. Anal. \textbf{57}, 291--323 (1974)

\bibitem{Korteweg1901}
Korteweg, D.J.: Sur la forme que prennent les equations du mouvement des
  fluides si l'on tient compte des forces capillaires.
\newblock Arch. Neerl. Sci. Ex. Nat. \textbf{6}, 1--24 (1901)

\bibitem{Maugin1980}
Maugin, G.A.: The method of virtual power in continuum mechanics: applications
  to coupled fields.
\newblock Acta Mech. \textbf{35}, 1--70 (1980)

\bibitem{Noll1990}
Noll, W., Virga, E.G.: On edge interactions and surface tension.
\newblock Arch. Rational Mech. Anal. \textbf{111}(1), 1--31 (1990)

\bibitem{Podio-Guidugli2002}
Podio-Guidugli, P.: Contact interactions, stress, and material symmetry, for
  nonsimple elastic materials.
\newblock Theoretical and Applied Mechanics \textbf{28-29}, 261--276 (2002)

\bibitem{Podio-Guidugli2007}
Podio-Guidugli, P.: On the aggregation state of simple materials.
\newblock In: \v{S}ilhav\'{y} (ed.) Mathematical modeling of bodies with
  complicated bulk and boundary behavior, \emph{Quaderni di Matematica},
  vol.~20, pp. 159--168 (2007)

\bibitem{Podio-Guidugli2009}
Podio-Guidugli, P., Vianello, M.: Fluidity and solidity notions for complex
  materials. (2009).
\newblock Unpublished.

\bibitem{Toupin1962}
Toupin, R.A.: Elastic materials with couple stresses.
\newblock Arch. Rational Mech. Anal. \textbf{11}, 385--414 (1962)

\bibitem{Toupin1964}
Toupin, R.A.: Theories of elasticity with couple-stresses.
\newblock Arch. Rational Mech. Anal. \textbf{17}, 85--112 (1964)

\end{thebibliography}

\end{document}